\documentclass[aip,jap,reprint,frontmatterverbose]{revtex4-1}
\usepackage{graphicx}
\usepackage{ulem}
\usepackage{amssymb}
\usepackage{amsmath}
\usepackage{todonotes}
\usepackage{wasysym}
\usepackage{relsize}
\usepackage{bm}
\bibpunct{}{}{,}{s}{}{}
\usepackage{float}
\usepackage{epstopdf}

\begin{document}

\title{Scaling in large area field emitters and the emission dimension}

\author{Rashbihari Rudra }
\affiliation{
Bhabha Atomic Research Centre,
Mumbai 400 085, INDIA}
\affiliation{
Homi Bhabha National Institute, Mumbai 400 094, INDIA}
\author{Debabrata Biswas}\email{dbiswas@barc.gov.in}
\affiliation{
Bhabha Atomic Research Centre,
Mumbai 400 085, INDIA}
\affiliation{
  Homi Bhabha National Institute, Mumbai 400 094, INDIA}

\begin{abstract}
  Electrostatic shielding is an important consideration for large area field emitters (LAFE) and results in a distribution of field enhancement factors even when the constituent emitters are identical. Ideally, the mean and variance together with the nature of the distribution should characterize a LAFE. In practice however, it is generally characterized by an effective field enhancement factor obtained from a linear fit to a Fowler-Nordheim plot of the $\text{I-V}$ data.  An alternate characterization is proposed here based on the observation that for a dense packing of emitters, shielding is large and LAFE emission occurs largely from the periphery, while well separated emitter tips show a more uniform or 2-dimensional emission. This observation naturally leads to the question of the existence of an emission-dimension, $D_e$ for characterizing LAFEs. We show here that the number of patches of size $L_P$ in the ON-state (above average emission) scales as $N(L_P) \sim L_P^{-D_e}$ in a given LAFE. The exponent $D_e$ is found to depend on the applied field (or voltage) and approaches $D_e = 2$ asymptotically. 
  \end{abstract}

\maketitle

\newcommand{\be}{\begin{equation}}
\newcommand{\ee}{\end{equation}}
\newcommand{\bea}{\begin{eqnarray}}
\newcommand{\eea}{\end{eqnarray}}
\newcommand{\Tbar}{{\bar{T}}}
\newcommand{\En}{{\cal E}}
\newcommand{\K}{{\cal K}}
\newcommand{\GC}{{\cal \tt G}}
\newcommand{\Lop}{{\cal L}}
\newcommand{\DB}[1]{\marginpar{\footnotesize DB: #1}}
\newcommand{\q}{\vec{q}}
\newcommand{\kt}{\tilde{k}}
\newcommand{\Lopn}{\tilde{\Lop}}
\newcommand{\noi}{\noindent}
\newcommand{\ovn}{\bar{n}}
\newcommand{\ovx}{\bar{x}}
\newcommand{\ovE}{\bar{E}}
\newcommand{\ovV}{\bar{V}}
\newcommand{\ovU}{\bar{U}}
\newcommand{\ovJ}{\bar{J}}
\newcommand{\calE}{{\cal E}}
\newcommand{\ovphi}{\bar{\phi}}
\newcommand{\zt}{\tilde{z}}
\newcommand{\rt}{\tilde{\rho}}
\newcommand{\tth}{\tilde{\theta}}
\newcommand{\nuv}{{\rm v}}
\newcommand{\ck}{{\cal K}}
\newcommand{\cc}{{\cal C}}
\newcommand{\ca}{{\cal A}}
\newcommand{\cb}{{\cal B}}
\newcommand{\cg}{{\cal G}}
\newcommand{\ce}{{\cal E}}
\newcommand{\cn}{{\cal N}}
\newcommand{\fn}{{\small {\rm  FN}}}
\newcommand\norm[1]{\left\lVert#1\right\rVert}



\section{Introduction}
\label{sec:intro}

A large area field emitter (LAFE) holds much promise as a cold source of electrons\cite{spindt68,spindt76,teo,dams2012,basu2015,parmee,whaley2018}. A typical LAFE consists of several thousands of individual field emitters packed together in a finite area $\ca$. The current density from a LAFE  is limited by two counter-acting effects. An increase in the number of emitters in $\ca$ results in more emitting tips. This however results in enhanced shielding between individual emitters which reduces the local field on individual tips. Thus, the number of emitters can be increased by packing more of them but beyond a point, this is counter-productive as individual contributions reduce sharply on account of shielding\cite{db_rudra}. To further complicate matters, the ideal packing density itself depends on the applied field and even the distance between the tip and the anode\cite{harris15,harris16,db_fef,db_rudra,db_anode,rudra_db_2019,db_rudra_2020,db_hybrid_2020,assis20}.

As shielding is non-uniform in any finite-sized LAFE, whether ordered or random, the field enhancement factor, $\gamma$, differs from tip to tip even if all emitters are identical in all respects. Thus, there exists a distribution $f(\gamma)$ of enhancement factors $\gamma$  at a given packing density (emitters per unit area) resulting in some interesting behavior. As a thumb rule, the periphery of a LAFE generally suffers minimal shielding and contributes more to the net current especially at higher packing densities while at higher applied fields or lower packing densities, even the emitters that are more centrally-located contribute to the current and start becoming visible in a current heat map.  

The complexity of a LAFE leads to difficulties in its characterization. Unlike a single emitter where the apex field enhancement factor, $\gamma$ and the apex radius $R_a$ are in-principle sufficient to determine the $\text{I-V}$ characteristics, a LAFE is conventionally characterized by an effective field enhancement factor and the notional emission area. In most instances however, the FN-plot is non-linear and the effective enhancement factor, $\gamma_{c}$, obtained from the slope of the regression line fitted to the experimental FN plot, is a poor representation of the LAFE.
Instead, the low and high field fits can be used to construct a 2-emitter class model with distinct enhancement-factors\cite{al-tabbakh,popov2020}. There is much however that needs to be understood since we are really dealing with a distribution of enhancement factors in a typical LAFE \cite{read2004,db_rudra,filippov2019}

Even in case of a single emitter, the pre-exponential factor in the empirical
expression for net current\cite{forbes2008}

\be
I_S = A_s E_0^{k_s} e^{-B_s/E_0} 
\ee

\noi
is not entirely a settled issue. For a flat emitter of area $A$, $I_s = A J_{MG}$, where $J_{MG}$ is the
Murphy-Good current density\cite{FN,Nordheim,burgess,murphy,jensen2003,forbes2006,FD2007,DF2008,jensen_book}.
Thus, $k_s = 2 - \nu$ where $\nu = \eta/6$ with
$\eta \approx 9.836~\text{(eV)}^{1/2} \phi^{-1/2}$.
When, the emitter is curved and the local field varies on its surface, $I_S = \int_A J_{MG} dA$.
For generic smooth endcap shapes\cite{db_ultram,physE}, the integration can be performed and
it is known\cite{db_dist}  that
$k_s \approx 3 - \nu$. There are exceptions however, most notably for the hemisphere on a cylindrical
post model. The value of $k_s$ has an important bearing on the experimental characterization of
single emitter tips, especially the notional emission area defined as $I_S/J_{MG}^{\text{apex}}$. Thus, while
the apex enhancement factor of a single tip is largely unaffected by the choice of $k_s$ and
can be determined from the slope of an FN-plot, the emission area
depends on the choice of $k_s$.

A LAFE consists of a collection of single emitters, each with a distinct apex field enhancement factor.
The net LAFE current $I_L$, can thus be expressed as

\be
I_L = \sum_{i=1}^{N} A_s^i E_0^{k_s} e^{-B_S^i /E_0} = E_0^{k_s} \left( \sum_{i=1}^{N} A_s^i e^{-B_S^i/E_0} \right)
\ee

\noi
assuming that they have identical shapes. The terms in the bracket distinguishes a LAFE from a single
emitter. If all the $N$ emitters are identical and well separated, $A_S^i$ and $B_S^i$ would be the same
for all emitters so that the net current $I_L \approx I_S N = I_S~ \rho \ca$ where $\rho$ is the number of
emitters per unit area and $\ca$ is the geometric area of the LAFE.

If the LAFE is not as sparse, $A_S^i$ and $B_S^i$ may be distinct for each emitter and it is likely that in
writing\cite{forbes2009}

\be
I_L = A_L E_0^{k_L} e^{-B_L/E_0} \label{eq:LAFE}
\ee

\noi
as in the single emitter case, $k_L \neq k_s$. As a matter of fact, it is not
apparent that the slope $B_L$ in an FN plot is directly related to the
enhancement factor of a LAFE even though it is commonly
used to extract the characteristic enhancement factor(s). There are thus
additional unresolved issues in dealing with a LAFE.

We propose here a markedly different approach to LAFE characterization based on the observation
that the glow pattern of a LAFE (the heat map) can vary from seemingly 1-dimensional peripheral
emission at low applied fields or  high packing densities, to the more uniform
seemingly 2-dimensional emission
at higher fields or lower packing densities. The question that we therefore address is whether
there exists any scaling behavior in a given LAFE that can capture the essence of the
glow pattern typical of a LAFE. More specifically, we wish to investigate whether the number
of patches (or covers) of size $L_P$ that outshines (ON state) the average behavior in a given LAFE,
scales as $L_P^{-D_e}$. If such
a relationship does indeed hold, it is also of interest to determine how the exponent
$D_e$ varies with $E_0$.

The information contained in the emission dimension can be useful in various ways
and can complement the $\text{I-V}$ data. 
It can for instance indicate the optimal emitter density and operating voltages that
can lead to uniform emission or even serve as a guide in designing a device.

The paper is organized as follows.
In Section \ref{sec:simulation}, we shall outline the methodology used for the scaling study
including a brief sketch of LAFE simulation. This is followed by the results
on scaling and finally a summary with a brief discussion on the experimental realization of the
scaling exponent.

\section{Simulation Methodology}
\label{sec:simulation}

Simulation of a LAFE having thousands of randomly placed individual emitters is central to the scaling
studies that we wish to perform. Clearly, such a massive task cannot be performed using
`exact' numerical methods such as finite element or boundary element techniques as it would
require enormous resources. An alternate approximate technique that is now well-tested,
is based on the line charge model. It involves the linear line charge
density and is applicable to hemi-ellipsoidal emitters. It is reasonably accurate
when the spacing between emitters is not too small compared to its height. A
generalization for other emitter shapes involving nonlinear line charge density
leads to the hybrid model for simulation of a general LAFE. We shall hereafter limit
our discussion to hemi-ellipsoidal emitters without any loss of generality.

\subsection{Current from a collection of emitters}

Consider a large area field emitter comprising of $N$ identical hemi-ellipsoidal
shaped emitters, each placed at ($x_i,y_i$), $i = 1,N$. 
The apex field enhancement factor $\gamma$ is defined as the ratio of the local
field at the apex, $E_a$ and the applied or macroscopic field $E_0$, i.e. $\gamma = E_a/E_0$.
It is a geometric quantity and depends principally on the ratio of its height $h$ and
the apex radius of curvature, $R_a$\cite{edgcombe2002,forbes2003,db_fef}.
For a collection of $N$ emitters, the field enhancement depends on the degree of
shielding and the proximity of the anode and a comprehensive modular theory has been
developed which provides an approximate value of the enhancement 
of an $i^{th}$ emitter in the
LAFE\cite{db_fef,db_anode,db_rudra,rudra_db_2019,db_rudra_2020,db_hybrid_2020}.
If the anode is considered to be far away, shielding effects dominate and the
apex field enhancement at the $i^{th}$ emitter is given by \cite{db_rudra},

\be
\gamma_i  \simeq  \frac{2h/R_a}{\ln\big(4h/R_a\big) - 2 + \alpha_{S_i}}  \label{eq:gamN0}
\ee

\noi
where $\alpha_{S_i} = \sum_{j\ne i} (\lambda_j/\lambda_i) \alpha_{S_{ij}} \simeq \sum_{j\ne i}  \alpha_{S_{ij}}$  and

\be
\alpha_{S_{ij}}  =  \frac{1}{\delta_{ij}}\Big[1 - \sqrt{1 + 4\delta_{ij}^2} \Big] + \ln\Big|\sqrt{1 + 4\delta_{ij}^2} + 2\delta_{ij} \Big| \nonumber
\ee

\noi
with $\delta_{ij} = h/\rho_{ij}$, $\rho_{ij} = [(x_i - x_j)^2 + (y_i - y_j)^2]^{1/2}$ being the distance between the $i^{th}$ and $j^{th}$ emitter on the cathode plane. In the above, $\lambda$ is the slope of the line charge density $\Lambda(z)$ (i.e. $\Lambda(z) = \lambda z$), obtained by projecting the surface charge density along the emitter axis\cite{jap2016}. The approximation $\lambda_i/\lambda_j \approx 1$ is found to be reasonable so long as the pair of emitters are not too close compared to their height.
Note that under this approximation, the shielding factor, $\alpha_{S_i}$ is a purely geometric quantity.
The predictions of Eq.~(\ref{eq:gamN0}) have been well tested\cite{rudra_db_2019} and found to
be accurate if the emitters are not too close to each other.

It is clear that if the emitter locations are randomly distributed, \{$\alpha_{S_i}$\} 
and hence \{$\gamma_i$\} are distinct. There is thus a distribution of enhancement factors
which can be determined on evaluating \{$\gamma_i$\} using Eq.~(\ref{eq:gamN0}).
If the mean inter-pin separation
$c$ is smaller than the height $h$ of the emitters, the distribution is skewed to the right
and emission is generally observed from the periphery. On the other hand, when $c > 2.5h$,
shielding has negligible effect on the local field enhancement.

The field enhancement factors \{$\gamma_i$\} together with the apex radius of curvature
$R_a$ can be used to determine the total LAFE current, $I_L$ as\cite{db_dist}

\be
I_L \approx \sum_{i=1}^N 2\pi R_a^2 g_i J_{MG}^i  \label{eq:lafeI}
\ee

\noi
where the area factor $g_i$ is

\be
g_i = \frac{\gamma_i E_0}{B_{\text{FN}} \phi^{3/2}} \frac{1}{(1-f_i/6)}
\ee

\noi
and the current density $J_{MG}$ is\cite{murphy,forbes2006}

\be
J_{MG}^i = \frac{A_{\text{FN}}}{\phi} \frac{(\gamma_i E_0)^2}{t_F^{2}} \exp\left(-\nu_F^i B_{\text{FN}} \phi^{3/2}/(E_0 \gamma_i)\right)
\ee

\noi
where, $A_{\text{FN}} = 1.541434 \times 10^{-6}$ $\text{A~eV~V}^{-2}$ and $B_{\text{FN}} = 6.830890~ \text{eV}^{-3/2} \text{V nm}^{-1}$ are the first and second Fowler-Nordheim constants, $\phi$ is the local work function of the emitting surface, $\nu_F^i = 1 - f_i + (f_i/6)\ln(f_i)$, $t_F = 1 + f_i/9 - (f_i/18)\ln(f_i)$ and $f_i \approx 1.44 \gamma_i E_0/\phi^2$.

Eq.~(\ref{eq:lafeI}) can be used to determine the current from a collection of N-emitters. It is
particularly suited for the scaling study which requires determination of the current
from a patch of size $L_P$ from any part of the LAFE. Throughout this study, we shall consider
the work-function $\phi$ to be uniform over the tip having a value $4.5$eV. It is
assumed that the work function remains constant throughout the simulation or experiment as the case may be.

\subsection{The scaling methodology}

Consider a LAFE of an arbitrary shape having an area $\ca$. Geometrically, it can be covered by $N_P$
non-overlapping square patches of size $L_P$. Since the LAFE itself is a 2-dimensional structure,
$N_P \sim L_P^{-2}$. This can be seen explicitly by writing $N_P = \ca/L_P^2$. Thus, we are clearly not
dealing with a geometric fractal.

Each patch covering the LAFE carries a net current $I_P$. Since the patches are non-overlapping,
$I_L = \sum I_P$. The average current density of the LAFE is $J_L = I_L/\ca$ and this serves
as a useful parameter in deciding whether a patch is emitting more than the background (ON state)
or less (OFF state). Thus, if $I_{P_k} \geq \alpha J_L L_P^2$, the $k^{th}$ patch is considered
to be in the ON-state and
assigned a value $S_k = 1$. Conversely, if $I_{P_k} < \alpha J_L L_P^2$, $S_k = 0$. We choose $\alpha$
to be unity.

The total number of patches of size $L_P$ in the ON-state is thus $N_{\text{ON}}(L_P) = \sum_k S_k$.
If the entire LAFE glows and $L_P$ is not too large, it is expected that $N_{\text{ON}}(L_P) \sim L_P^{-2}$.
On the other hand, if the LAFE glows at the periphery, then  $N_{\text{ON}}(L_P) \sim L_P^{-1}$ provided
$L_P$ is small enough that bulk and surface effects can be distinguished.

Clearly, if $L_P$ is too small, statistical errors are likely to be large,
especially in the low field regime where a patch with only a few emitters
may struggle to be in the ON state. On the other hand, if
$L_P$ is comparable to the size of the LAFE, saturation effects can set in.
Thus, if a scaling exists, it must be in the intermediate region of patch size $L_P$.
For convenience and comparison, the patch size $L_P$ will be considered as integer
multiples of the mean separation $c$ i.e. $L_P = c N_C$. Since the mean separation is $c$,
the number of emitters in the patch of size $L_P$ is the area of the patch $L_P^2$ divided by
the average area occupied by a single emitter, $c^2$. Thus the number of emitters in a patch
of size $L_P$ is $L_P^2/c^2 = N_C^2$.

\section{Scaling Results}

A typical random LAFE simulation starts with a random
generation of $N$ pin locations \{$(x_i,y_i)$\} distributed on a given shape 
having area $Nc^2$, using a uniform random number generator. This is followed
by evaluation of \{$\alpha_{S_i}$\} and \{$\gamma_i$\} using Eq.~(\ref{eq:gamN0}). Once the individual
enhancement factors are evaluated, the current from each pin can be
determined and used to compute the total current, the average current density
the current from a given ($k^{th}$) patch of size $L_P$ and the corresponding state $S_k$.

In the following, we shall consider LAFEs with hemi-ellipsoidal emitter pins of height $h = 1500\mu$m and
base radius of curvature $b = 12.5\mu$m. This implies an apex radius of curvature
$R_a = b^2/h \approx 104.17$nm\cite{curvature_effects,db_curvature,db_rr_jap2021}.
The number of emitters in the LAFE is typically $3.6\times 10^5$
and the mean separation $c$ varies from $1000\mu$m to $2500\mu$m. Since the number of pins is
held fixed, the LAFE area $\ca$ increases with $c$. Note that the dimensions considered
here are not necessarily the typical ones involved in field emission but serve to illustrate
the essential ideas of scaling.

\begin{figure}[htbp]
  \vskip -0.75cm
  \hspace*{-0.8cm}\includegraphics[width=.6\textwidth]{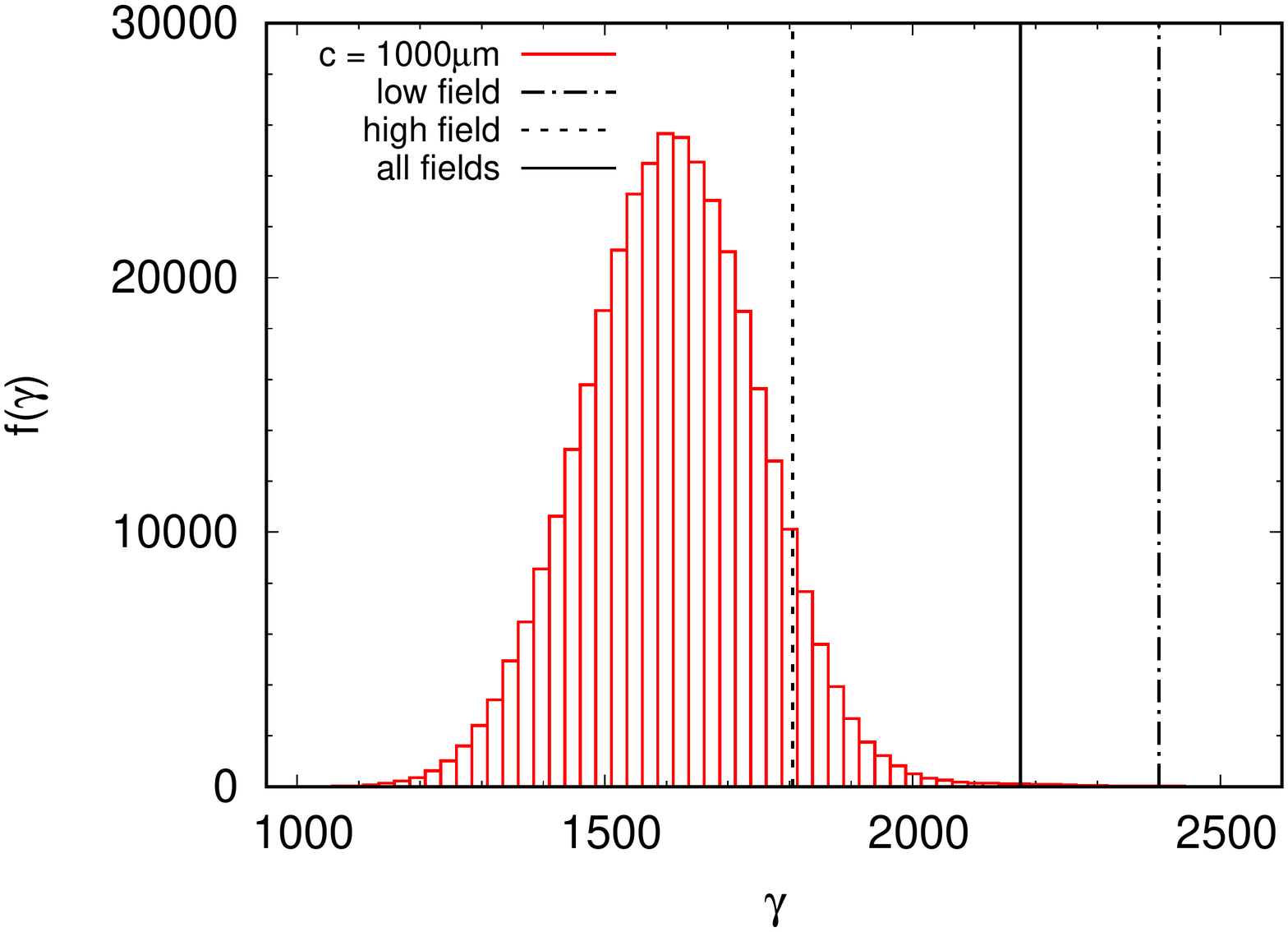}
  \vskip -0.75cm
\caption{The distribution of enhancement factors $f(\gamma)$ at $c = 1000\mu$m, $h = 1500\mu$m
  along with the characteristic values of $\gamma$ extracted from an FN plot.}
\label{fig:hist_1000}
\end{figure}

A typical frequency distribution of enhancement factors for $c=1000\mu$m and $h = 1500\mu$m is shown
in Fig.~\ref{fig:hist_1000}. The frequency distribution is obtained from the \{$\gamma_i$\}
evaluated using Eq.~(\ref{eq:gamN0}) which is based on the line charge model.
The vertical lines mark the extracted values of the
characteristic enhancement
factor from the FN plot ($\ln(I/E_0^2)$ vs $1/E_0$),
assuming that $B_L = B_{FN} \phi^{3/2}/\gamma_c$ where
$\gamma_c$ is the characteristic
enhancement factor in the range of applied fields considered for the fit. The
full range of applied field is $E_0 \in [0.5,3]$V/$\mu$m. Clearly, the low field and
high field values of $\gamma_c$ differ considerably and the FN-plot is nonlinear.

\begin{figure}[htbp]
 \vskip -1.0cm 
 \hspace*{-0.7cm}\includegraphics[width=.58\textwidth]{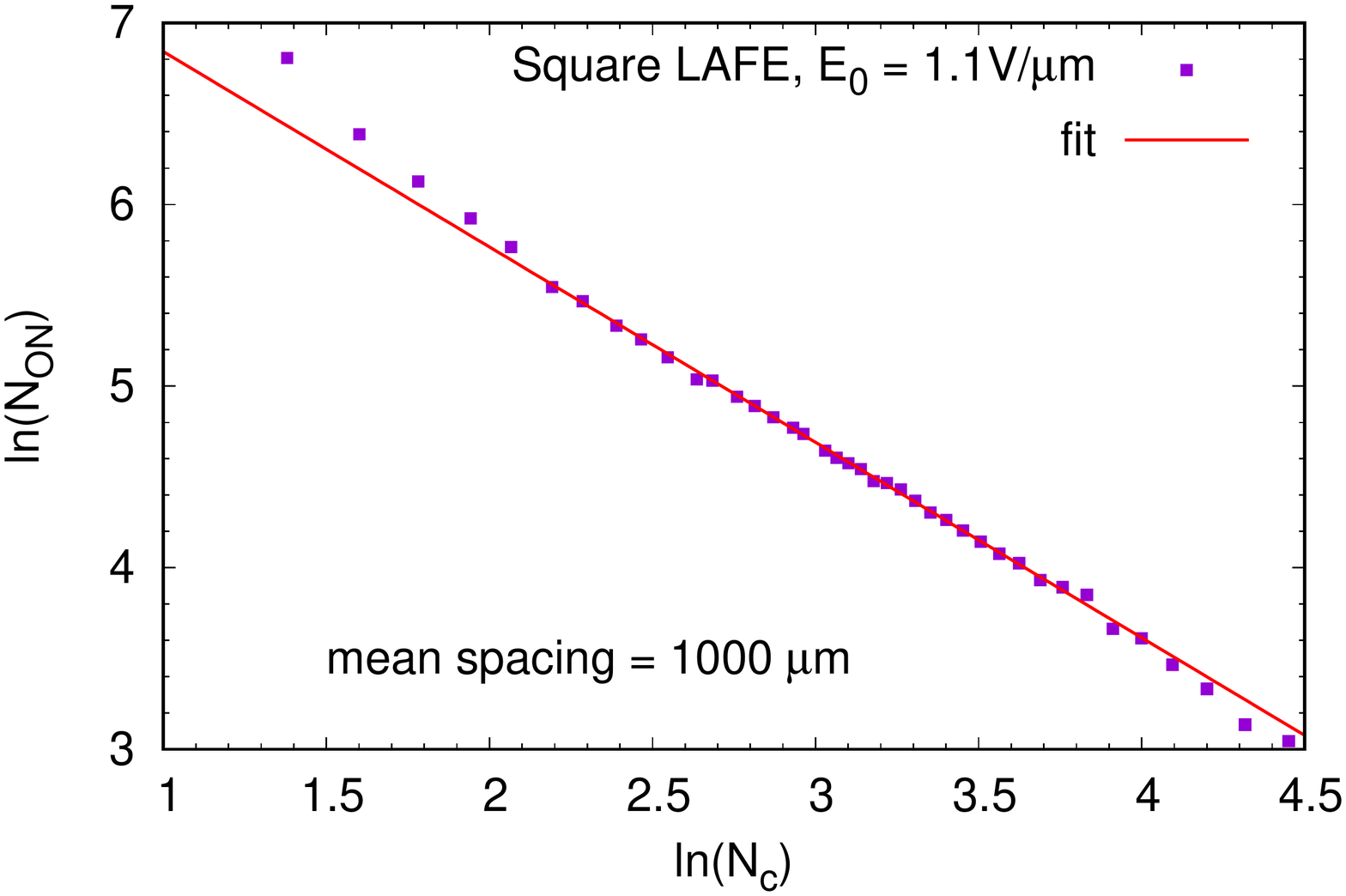}
 \vskip -0.85cm 
\caption{Scaling behavior for a square LAFE with mean spacing $c = 1000\mu$m.
  The number, $N_{ON}$, of patches in ON state scales as a power law with the patch length
  $L_P = cN_C$, as shown
by the straight line fit.}\label{fig:square_ipsep1000}
\end{figure}

The wide distribution of enhancement factors can lead to interesting local emission properties. 
Fig. ~\ref{fig:square_ipsep1000} shows a typical $\ln(N_{\text{ON}})$ vs $\ln(N_c)$ plot for average inter-pin separation $1000 \mu$m  at an applied field $E_0 = 1.1 \times 10^{-3}$ V/nm for a square LAFE. For intermediate patch size, the fit to a straight line is very good clearly indicating a power law scaling $N_{ON} \sim  L_P^{-D_e}$ where $D_e$ is the slope of the fitted straight line. In this instance, $D_e \approx 1.07$, which is much smaller than 2. We shall hereafter
refer to the scaling exponent $D_e$ as the Emission-Dimension.

\begin{figure}[htbp]
  \vskip -0.65cm 
  \hspace*{-0.7cm}\includegraphics[width=.58\textwidth]{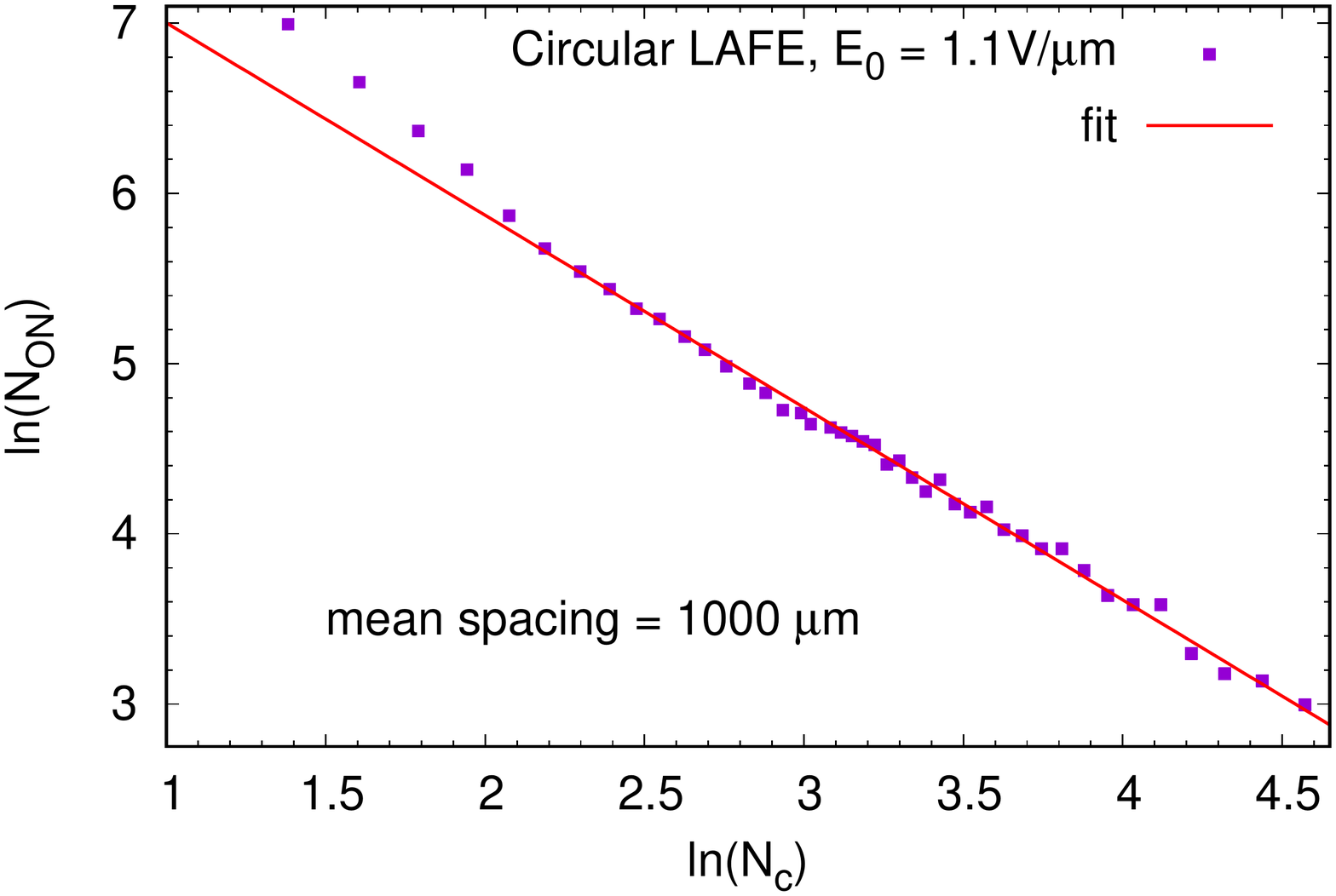}
  \vskip -0.65cm 
\caption{Scaling behavior for a circular LAFE with mean spacing $c = 1000\mu$m.
The slope of the straight line is approximately -1.12.}\label{fig:circle_ipsep1000}
\end{figure}

A similar behavior is observed for a Circular LAFE. The number of patches in the ON state
again scales as a power law for $E_0 = 1.1 \times 10^{-3}$ V/nm and $c = 1000\mu$m. In this
case, $D_e \approx 1.12$. Since the number of emitters in the square and circle LAFE are the
same, their net area is identical for a fixed $c$. Thus, if the area is $L^2$, 
the ratio of perimeter to area is $4/L$ for a square LAFE. For a circle having area $L^2$,
its radius is $L/\sqrt{\pi}$
and the ratio of its circumference and area is $(2\pi L/\sqrt{\pi})/L^2 = 2\sqrt{\pi}/L < 4/L$.
Thus, the square LAFE is more likely to have a peripheral glow pattern and hence the emission
dimension is likely to be smaller.
Note that for both the square and circular LAFE at $E_0 = 1.1~\text{V/}\mu$m,
the fit is in the range $N_C \in [8,90]$ corresponding to $\ln(N_C) \in [2,4.5]$.

\begin{figure}[htbp]
 \vskip -0.75cm 
 \hspace*{-0.7cm}\includegraphics[width=.58\textwidth]{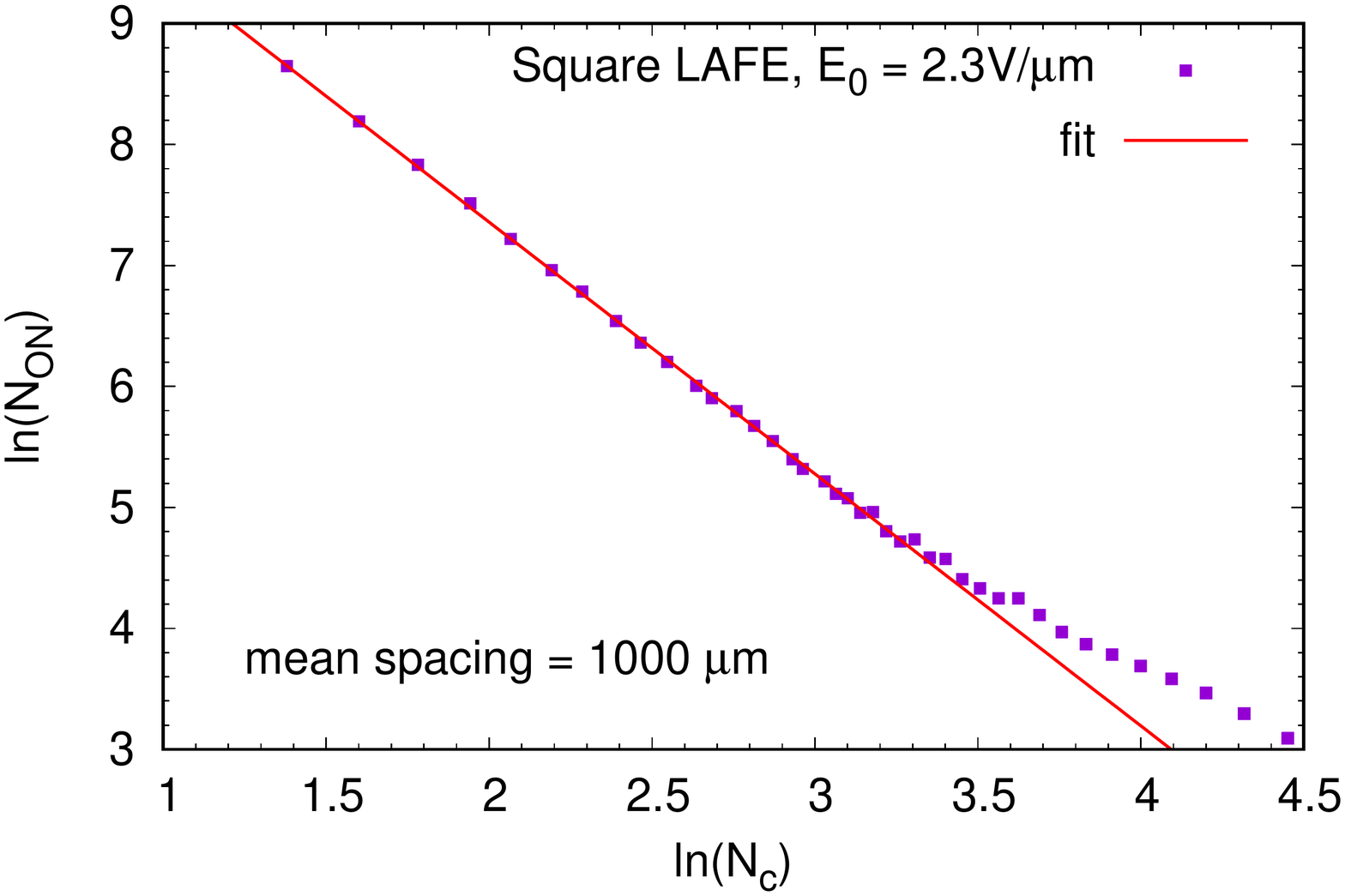}
 \vskip -0.75cm 
 \caption{Scaling behavior for a square LAFE at a higher applied field ($E_0 = 2.3$V/$\mu$m).}
   \label{fig:square_higherE0}
\end{figure}

The  Emission-Dimension, $D_e$, should depend on
the applied field $E_0$ as well as the mean separation $c$ between the emitters.
Note that with an increase in applied field $E_0$, the small patch-size
contributes in a manner similar to the intermediate patch size as seen in
Fig.~\ref{fig:square_higherE0} where the straight line fit extends over
$N_C \in [4,36]$ corresponding to $\ln(N_C) \in [1.38,3.6]$.
For purposes of determining the variation of $D_e$ with $E_0$, we consider as a thumb rule
a uniform intermediate range starting from $\ln(N_C) \approx 2.2$ to $\ln(N_C) \simeq 3.6$
corresponding to $N_C \in [9,36]$.

\begin{figure}[htbp]
  \vskip -0.75cm 
  \hspace*{-0.7cm}\includegraphics[width=.58\textwidth]{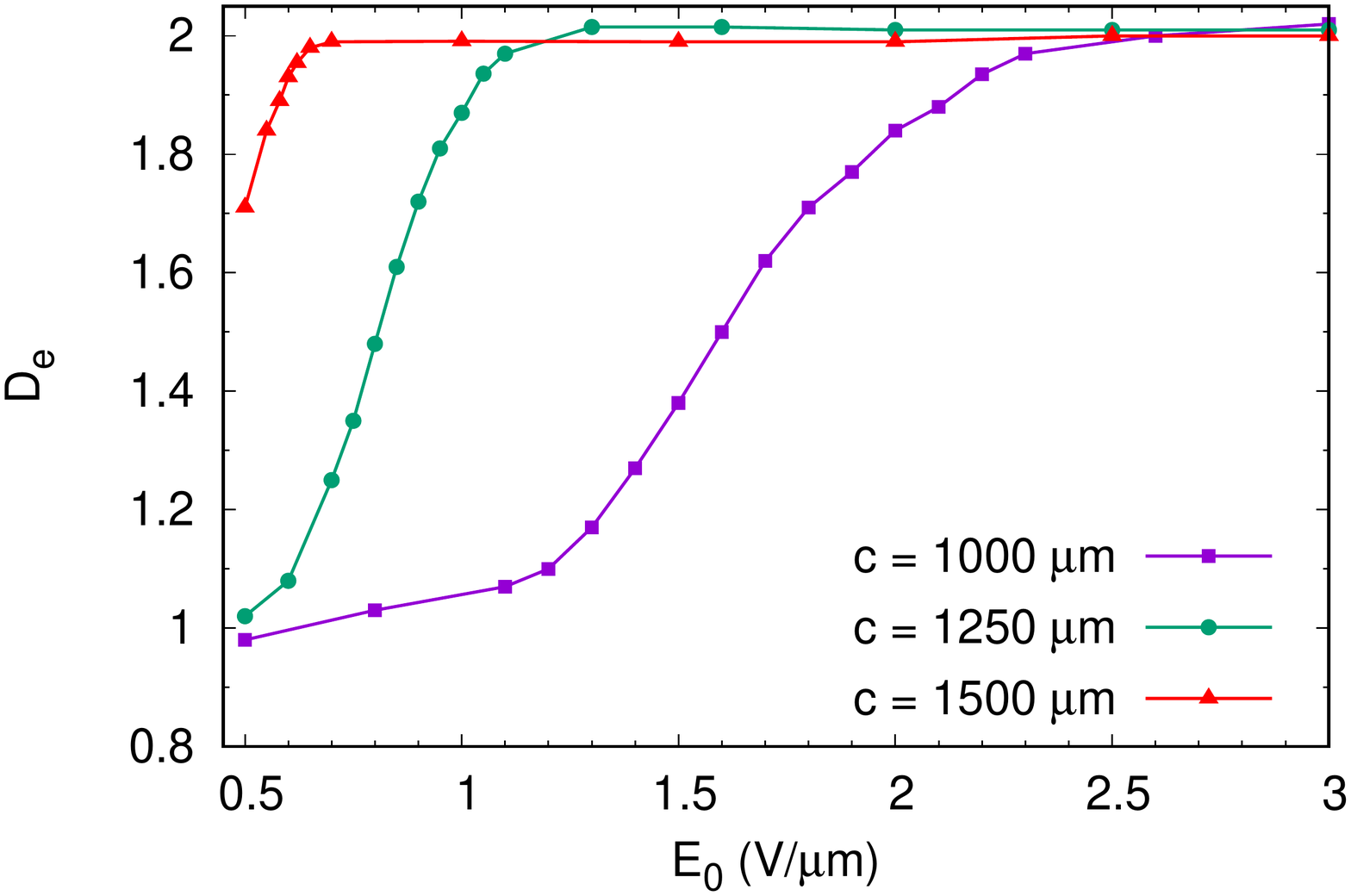}
  \vskip -0.75cm 
  \caption{Variation of the emission dimension $D_e$ with applied electric field
    $E_0$ for three separate mean separation ($c$) of emitters.}
\label{fig:D_vs_E}
\end{figure}

\begin{figure}[htbp]
  \vskip -0.75cm
  \hspace*{-0.8cm}\includegraphics[width=.6\textwidth]{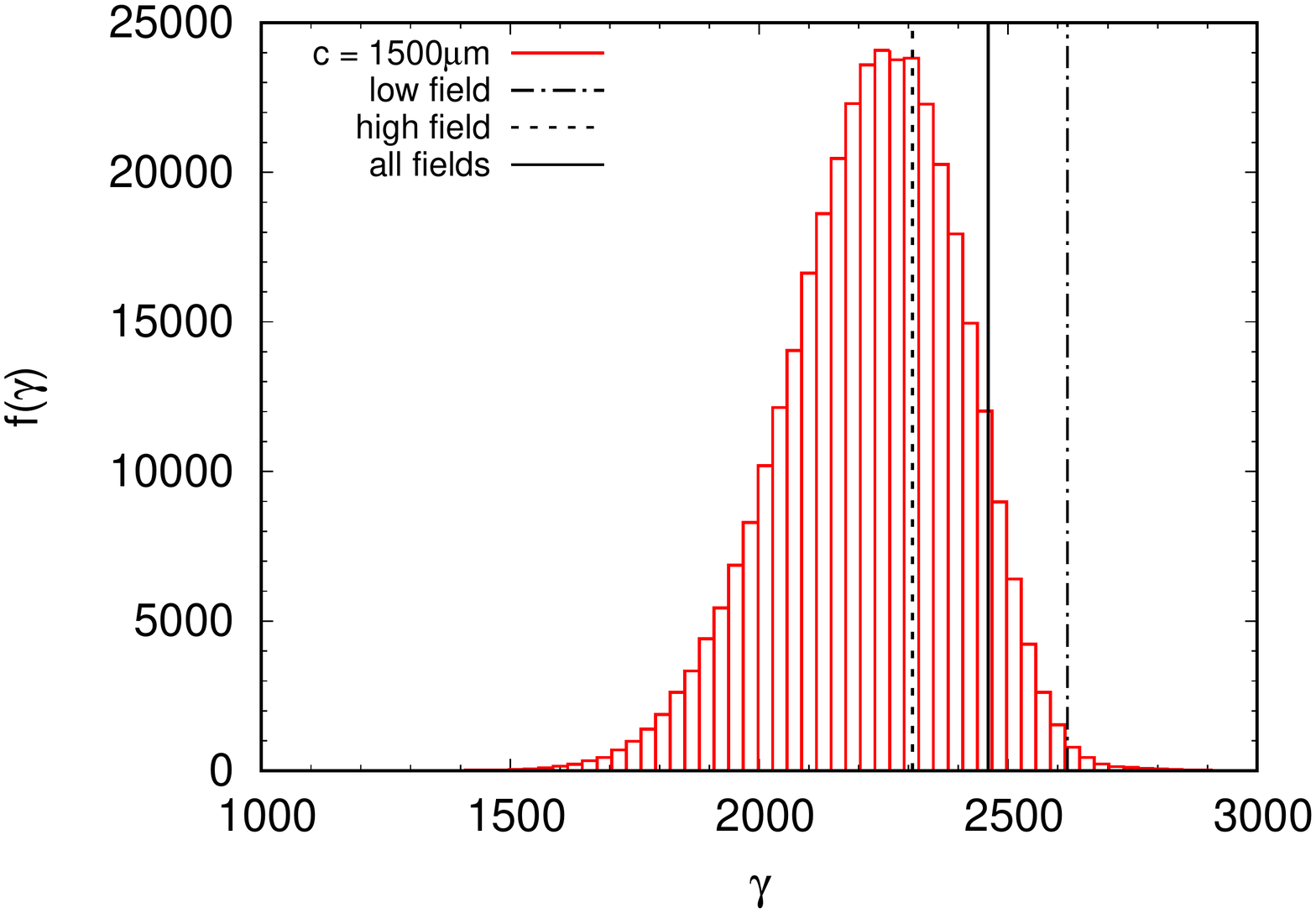}
  \vskip -0.75cm
\caption{The distribution of enhancement factors $f(\gamma)$ at $c = 1500\mu$m, $h = 1500\mu$m
  along with the characteristic values of $\gamma$ extracted from an FN plot.}
\label{fig:hist_1500}
\end{figure}

Fig.~\ref{fig:D_vs_E} shows the variation of the emission dimension $D_e$ with
the applied field $E_0$ for three different LAFEs, having mean separation
$c$ equal to $1000\mu$m, $1250\mu$m, and $1500\mu$m respectively. Clearly, $D_e \rightarrow 2$
for large $E_0$ in all cases since  the interior of the LAFE starts contributing
substantially to the net emission current. A similar trend can be observed as the
mean separation increases. For instance at $E_0 = 1$V/$\mu$m, $D_e$ increases from about 1
to 1.8 as the mean separation $c$ increases from $1000\mu$m to $1250\mu$m. For
$c = 1500\mu$m, $D_e \approx 2.0$.
The rapid change is due to the sharp rise in the field enhancement factor with $c$ and
the change in $f(\gamma)$ from being right-skewed to an almost symmetric distribution
at $c = 1500\mu$m as seen in Fig.~\ref{fig:hist_1500}.

Note that the extracted values of $\gamma_c$ at low fields, high fields and the full range of applied
field, are now more representative of the distribution. Moreover, the relative gap between the
low and high field values is now smaller and as $c$ increases further, the three values come closer
and stand near the peak (mode). This also results in a linear FN-plot for the same range of
applied fields and coincides with the emission dimension being $D_e  = 2$.

\section{Discussion and summary}

The study presented in this paper clearly establishes scaling properties of a
large area field emitter and the existence of an emission dimension $D_e$ that
depends on the applied field. It is an especially useful characterization when
emitters are tightly packed such that the mean spacing is smaller than the height
of the emitter. This results in a large range of variation in 
$D_e$ with the applied electric field. 

While the theoretical study presented here was based on vertically-standing
emitters of identical height, the idea of scaling may be more generally applicable
in large area field emitters and can be investigated both theoretically and (largely)
experimentally.

The need to understand and characterize LAFEs beyond the conventional $\text{I-V}$
analysis was motivated in part by recent experimental progress in studying local
properties of a LAFE \cite{read2004,filippov2019,popov2018,popov2020a}.
We believe, an equivalent experimental study of
scaling in a large area field emitters is possible and will present us with
a greater understanding of their emission properties and serve as a guide in
designing practical devices.

\vskip 1.0cm
{\it Acknowledgements}: The scaling study was taken up following a remark by Prof. G. Ravikumar on the emission
properties of LAFE. The authors also acknowledge fruitful discussions with Dr. Raghwendra Kumar.

\vskip 1.0cm
{ \it Data Availability}:  The data that supports the findings of this study are available within the article.

\vskip 0.05 in


\section{Reference}


\begin{thebibliography}{99}
\bibitem{spindt68} C. A. Spindt, J. Appl. Phys. 39, 3504 (1968).
\bibitem{spindt76} C. A. Spindt, I. Brodie, L. Humphrey, and E. R. Westerberg, J. Appl. Phys. 47, 5248 (1976).
\bibitem{teo}  K.~B.~K.~Teo, E.~Minoux, L.~Hudanski, F.~Peauger, J.~P.~Schnell, L.~Gangloff, P.~Legagneux, D.~Dieumegard, G.~A.~J.~Amaratunga and W.~I.~Milne, Nature 437, 968 (2005).
\bibitem{dams2012} F. Dams, A. Navitski, C. Prommesberger, P. Serbun, C. Langer, G.
  Muller, and R. Schreiner, IEEE Trans. Electron Devices 59, 2832 (2012).
\bibitem{parmee} R. J. Parmee, C. M. Collins, W. I. Milne, and M. T. Cole, Nano Convergence 2, 1 (2015). 
\bibitem{basu2015} A. Basu, M. E. Swanwick, A. A. Fomani, and L. F. Vela\'{s}quez-Garcia, J. Phys. D: Appl. Phys. 48, 225501 (2015).
\bibitem{whaley2018} D. R. Whaley, C. Armstrong, C. E. Holland, C. A. Spindt, and P. R. Schwoebel, in 31st International Vacuum Nanoelectronics Conference (IVNC) (IEEE, New York, 2018), p. 1.
\bibitem{harris15} J.~R.~Harris, K.~L.~Jensen, D.~A.~Shiffler, AIP Adv. 5 (2015) 087182.
\bibitem{harris16} J.~R.~Harris, K.~L.~Jensen, W.~Tang, D.~A.~Shiffler, J. Vac. Sci. Technol. B 34 (2016) 041215.
\bibitem{db_fef} D.~Biswas, Phys. Plasmas 25, 043113 (2018).
\bibitem{db_rudra} D.~Biswas and R.~Rudra, Physics of Plasmas 25, 083105 (2018).
\bibitem{db_anode} D.~Biswas, Physics of Plasmas, 26, 073106 (2019).
\bibitem{rudra_db_2019} R.~Rudra and  D.~Biswas, AIP Advances, 9, 125207 (2019).
\bibitem{db_rudra_2020} D.~Biswas and R.~Rudra, J. Vac. Sci. Technol. B, 38, 023207 (2020).  
\bibitem{db_hybrid_2020} D.~Biswas,  J. Vac. Sci. Technol. B38, 063201 (2020).
\bibitem{assis20} T.~A. de Assis, F.~F.~Dall'Agnol, and M.~Cahay, Applied Physics Letters 116, 203103 (2020).
\bibitem{al-tabbakh} A.~A.~Al-Tabbakh, Ultramicroscopy, 218, 113087 (2020).
\bibitem{popov2020} E.~O. Popov,  A.~G. Kolosko, S.~V.~Filippov, T.~A. de Assis, Vacuum, 173, 109159 (2020).
\bibitem{read2004} F. H. Read and N. J. Bowring, Nucl. Instrum. Methods Phys. Res. A 519, 305 (2004).
\bibitem{filippov2019} S.~V.Filippov, A.~G.~Kolosko, R.~M.~Ryazanov, E.~P.~Kitsyuk, E.~O.~Popov, IOP Conference Series: Materials Science and Engineering 525, 012051 (2019).
\bibitem{forbes2008} R.~G. Forbes, Appl. Phys. Lett. 92, 193105 (2008).
\bibitem{FN} R.~H.~Fowler and L.~Nordheim, Proc. R. Soc. A 119, 173 (1928).
\bibitem{Nordheim} L.~Nordheim, Proc. R. Soc. A 121, 626 (1928).
\bibitem{burgess} R.~E.~Burgess, H.~Kroemer, J.~M.~Houston, Phys. Rev. 90, 515 (1953).
\bibitem{murphy} E.~L.~Murphy and R.~H.~Good, Phys. Rev. 102, 1464 (1956).
\bibitem{jensen2003} K.~L.~Jensen, J. Vac. Sci. Technol. B 21, 1528 (2003).
\bibitem{forbes2006} R.~G.~Forbes, App. Phys. Lett. 89, 113122 (2006).
\bibitem{FD2007} R.~G.~Forbes and J.~H.~B.~Deane, Proc. R. Soc. A 463, 2907 (2007).
\bibitem{DF2008} J.~H.~B.~Deane and R.~G.~Forbes, J. Phys. A: Math. Theor. 41, 395301 (2008).
\bibitem{jensen_book} K.~L.~Jensen, {\it Introduction to the physics of electron emission}, Chichester, U.K., Wiley, 2018.
\bibitem{db_ultram} D.~Biswas, G.~Singh, S.~G.~Sarkar and R.~Kumar, Ultramicroscopy 185, 1 (2018).
\bibitem{physE} D.~Biswas, G.~Singh and R.~Ramachandran, Physica E 109, 179 (2019).
\bibitem{db_dist} D.~Biswas, Physics of Plasmas 25, 043105 (2018).
\bibitem{forbes2009} R.~G. Forbes, J. Vac. Sci. Technol. B27, 1200 (2009).
\bibitem{edgcombe2002} C.~J.~Edgcombe, and U.~Valdr\`{e}, Philosophical Magazine B 82, 987 (2002).
\bibitem{forbes2003} R.~G.~Forbes, C.~J.~Edgcombe, and U.~Valdr\`{e}, Ultramicroscopy 95, 57 (2003).
\bibitem{jap2016} D.~Biswas, G.~Singh and R.~Kumar, J.~Appl.~Phys. 120, 124307 (2016).
\bibitem{curvature_effects} For $R_a < 100$nm, curvature effects in the field emission
  current density become prominent and a different treatment is required.
  See [\onlinecite{db_curvature,db_rr_jap2021}].
\bibitem{db_curvature} D.~Biswas and R.~Ramachandran, J. Vac. Sci. Technol. B 37, 021801 (2019).
\bibitem{db_rr_jap2021} D.~Biswas and R.~Ramachandran, J.~Appl.~Phys. 129, 194303 (2021).    
\bibitem{popov2018} E.~O. Popov, A.~G. Kolosko, S.~V. Filippov, and E.~I. Terukov,
  J. Vac. Sci. Technol. B36, 02C106 (2018).
\bibitem{popov2020a} E. O. Popov, A. G. Kolosko, S. V. Filippov, E. I. Terukov, R. M. Ryazanov, and E. P. Kitsyuk, J. Vac. Sci. Technol., B 38, 043203 (2020).
\end{thebibliography}
\end{document}